\documentclass[11pt,twoside]{article}

\usepackage{asp2006}
\usepackage{epsf}
\usepackage{psfig}
\usepackage{lscape}
\usepackage{natbib}
\usepackage{amssymb}

\newcommand\kms{{\rm\,km\,s^{-1}}}

\newcommand\ulsr{U_{\rm LSR}}
\newcommand\vlsr{V_{\rm LSR}}
\newcommand\wlsr{W_{\rm LSR}}

\begin{document}
\title{The nature of the metal-rich thick disk}   
\author{T. Bensby,$^{1}$ A.R. Zenn,$^{1}$ M.S. Oey,$^{1}$ and S. Feltzing$^{2}$}   
\affil{$^{1}$ Dept. of Astronomy, University of Michigan, 
830 Dennison Bldg, \\500 Church Street, Ann Arbor, MI 48109-1042, USA}
\affil{$^{2}$ Lund Observatory, Box 43, SE-22100 Lund, Sweden}

\begin{abstract} 
We have traced the Galactic thick disk to its,to date, highest metallicities.
Based on high-resolution spectroscopic observations of 187 F and G
dwarf stars that kinematically can be associated either with the thin disk
(60 stars) or with the thick disk (127 stars), we find that the thick disk 
stars reach at least solar metallicities, and maybe even higher. 
This finding is independent of the $\ulsr$, $\vlsr$ and $\wlsr$
velocities of the stars. 
\end{abstract}

\section{Introduction}

The metal-rich stars of the Galactic thick disk may be an
the evolutionary interface to the thin disk.  The metal-rich extreme
of the thick disk will help us to understand 
whether this transition is smooth and whether there was a hiatus in the star
formation history between the two disks. 
Having a well-established upper metallicity limit for the thick disk 
is also crucial for studies of possible age-metallicity relations in the 
thick disk. 

However, despite the efforts of several recent high-resolution
spectroscopic surveys of the Galactic thick disk, it is unclear what
its high-metallicity limit is.  For instance,
\cite{fuhrmann1998,fuhrmann2004} suggests that the thick 
disk stops at $\rm [Fe/H]\,\approx\,-0.3$; \cite{reddy2006} find that
the thick disk may not include stars with $\rm [Fe/H] > -0.3$;
and \cite{mishenina2004} suggest, from their data, that the star
formation in the thick disk stopped when the enrichment was
$\rm [Fe/H] = -0.3$. It should be noted that all of these studies
actually contain stars that are more metal-rich and that have
kinematic properties that could classify them as thick disk stars.
However, \cite{fuhrmann1998,fuhrmann2004} regards these stars as ``transition
objects" and it is unclear whether they should be treated as thin or thick disk
stars, or neither. \cite{reddy2006} find that metal-rich, potential thick
disk stars mainly follow thin disk trends \citep{reddy2003}, and
\cite{mishenina2004} claim that their metal-rich stars with
hot kinematics cannot be assigned to the thick disk due to both
their highly eccentric orbits, and their low Galactic vertical
distribution (low $Z_{\rm max}$). Because of this, they say
that the origin of these stars should be sought elsewhere, such as e.g.
in the Hercules stream \citep[see e.g.][]{famaey2005}.
However, even if potential Hercules stream stars are excluded
\cite{soubiran2005} find, in their compilation of high-resolution data
from the literature, a considerable number of stars
with thick-disk--like kinematics at high metallicities. But they
also conclude that their sample has too few metal-rich stars with 
thick disk kinematics to really verify their thick disk origin.

In our own studies of the thin and thick disks \citep{bensby2003,bensby2005}
we do not insist on an upper metallicity limit
for the thick disk, but assume that the stars we see with thick-disk--like
kinematics could be members, regardless of their
metallicities. And what we do find is that the kinematically hot stars
that we associate with thick disk, and that have $\rm [Fe/H]>-0.3$,
differ significantly from the thin disk stars at the same metallicity.
This applies both in terms of abundance ratios (e.g. [Mg/Fe]), as well as ages
\citep{bensby2003,bensby2004,bensby2005,bensby2006,feltzing2003,feltzing2006}.
Also, in \cite{bensby2007letter} we present a detailed abundance study of
60 F and G dwarf stars that kinematically are likely to be members of the
Hercules stream. What we find is that these stars do not form a genuine
population but that instead, they are likely to be a mixture of stars from the
thin and thick disk populations, verifying that their kinematic
properties are probably due to dynamical interactions with the Galactic bar.

As the current data on the metal-rich thick disk evidently is very
sparse and since there might be indications that the stars of the
Hercules stream could be a ``metal-rich thick disk" it is important to
observe and establish the extreme metal-rich limit of the thick disk.
We have therefore carried out an extensive spectroscopic
survey of metal-rich stars that kinematically can be associated with the
Galactic thick disk. Special care has been applied to exclude
stars that have space velocities typical of the Hercules stream.

Here, we will focus on two elements that
show distinct abundance trends for the thin and thick disks:
Mg \citep[e.g.,][]{fuhrmann2004,feltzing2003,bensby2003,bensby2005} and
Ba \citep[e.g.,][]{mashonkina2003,bensby2005}.
Other $\alpha$-elements, iron peak elements, and $r$- and $s$-process
elements will be presented in an upcoming paper
(Bensby et al., in prep.),
wherein we also will describe the observations, data reductions,
abundance analysis, and so forth.

\section{Stellar sample}

\begin{figure}
\plottwo{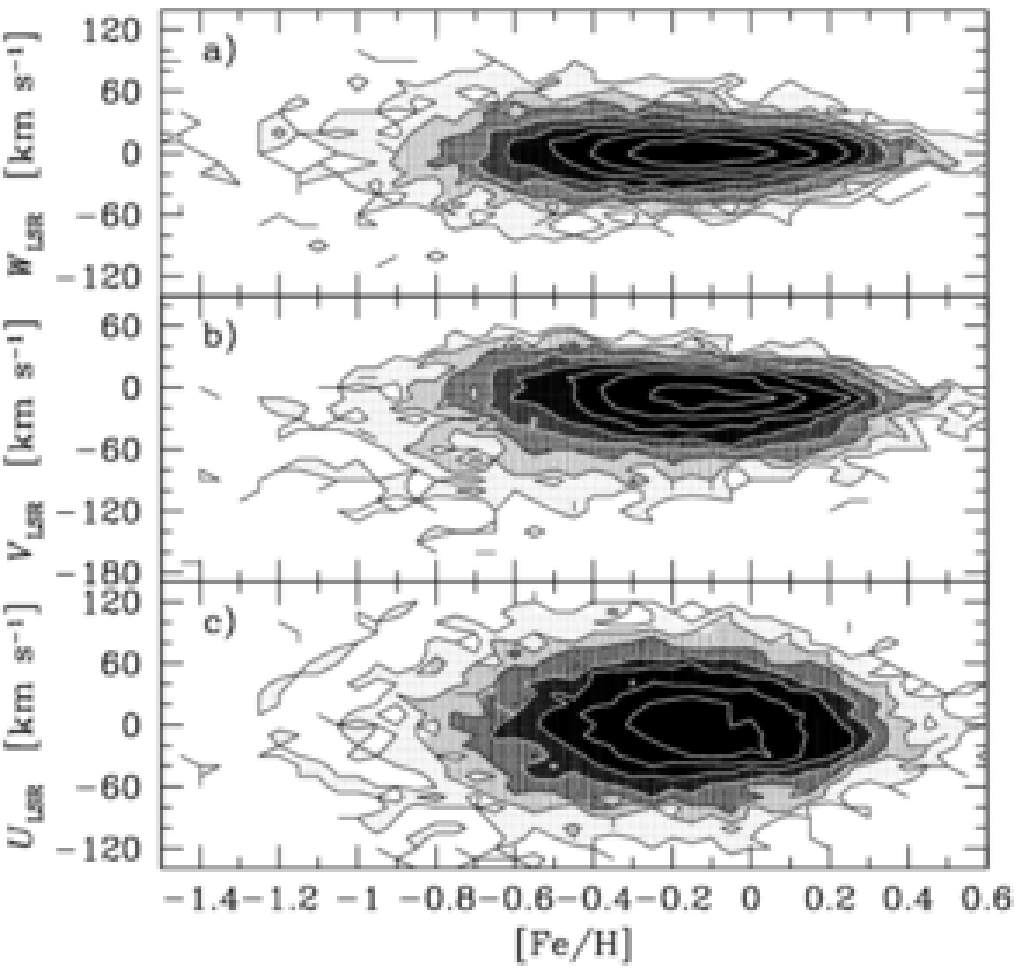}{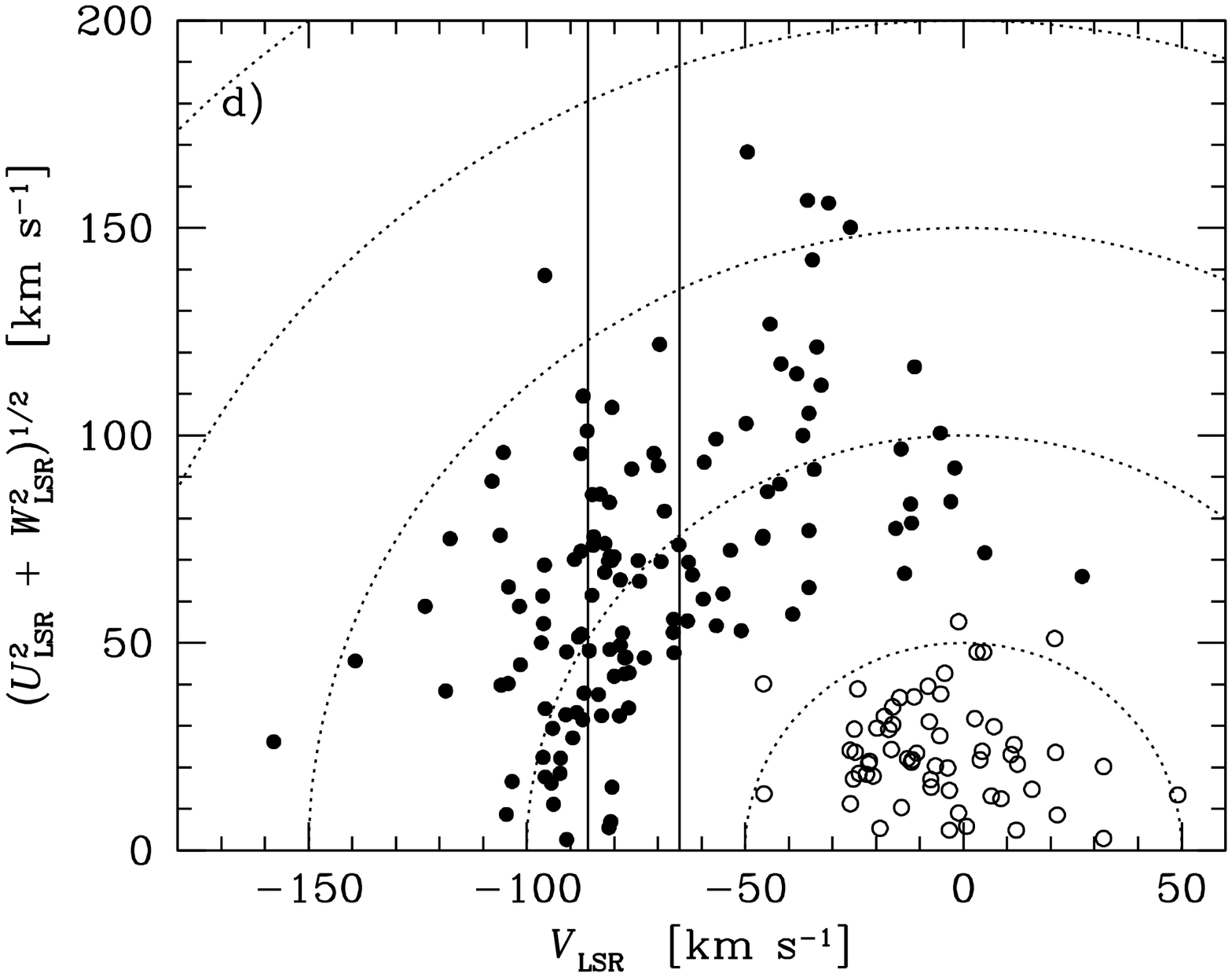}
\caption{
        {\bf a}\,--\,{\bf c:}
        Velocity-metallicity distribution for the $\sim 13\,000$ F and
        G dwarf stars in the \cite{nordstrom2004} catalogue.
        The contours are; black lines: 1, 3, 7, and white lines:
        15, 30, 60, 120, 240.
        {\bf d}: Toomre diagram for the stellar sample (including both the new 
	thick disk stars observed with MIKE as well as
	the thin
	and thick disk stars from \citealt{bensby2003,bensby2005}).
	Stars with thin and thick disk kinematics are marked by open and 
	filled circles, respectively.
        Dotted lines indicate constant total space velocities,
         $v_{\rm tot} = 
        (U^{2}_{\rm LSR} + V^{2}_{\rm LSR} + W^{2}_{\rm LSR})^{1/2}$,
        in steps of 50 km~s$^{-1}$.
	The two vertical lines at $\vlsr=-86\,\kms$ and $\vlsr=-65\,\kms$
	divides the thick disk sample into three equally sized 
	subsamples of 42, 45, and 42 stars (see 
	Sect.~\ref{sec:results}).
}
\label{fig:contour}
\end{figure}

What characterizes the thick disk is that its stars are
kinematically hot and rotationally lag behind the Local Standard
of Rest (LSR) by some $50 \kms$. In Figs.~\ref{fig:contour}a and b
we plot how the $\ulsr$, $\vlsr$ and $\wlsr$ velocities for the stars 
in the \cite{nordstrom2004} catalogue ($\sim13\,240$ F and G dwarf stars
with 3-dimensional kinematic information)
are distributed as a function of metallicity. Even though there might
be a slight decrease in the number of high-velocity stars as one goes to higher
metallicities, there is no sharp drop-off. Instead, there seems to
be a significant number of stars with both high metallicities and high
velocities.

We used the kinematic method from \cite{bensby2003, bensby2005}
to define samples of potential thin disk and thick disk stars.
This method assumes that a stellar population has a Gaussian velocity
distribution and constitutes a certain fraction of the stars in the solar
neighbourhood. Assuming that the solar neighbourhood is a mixture
of only the thin disk ($D$), the thick disk ($TD$), the Hercules stream ($Her$),
and the halo ($H$), we then calculate the probabilities for individual stars
(with known space velocities) to belong to any of the populations.
By forming probability ratios we select thick disk stars as those
that have probabilities of belonging to the thick disk that are at least
twice the probabilities of belonging to any of the other populations,
i.e. $TD/D>2$, $TD/H>2$, and $TD/Her>2$.

The 91 new potential metal-rich\footnote{The new thick disk sample presented
here only includes stars with $\rm [Fe/H]>-0.60$. Another sample of new 
thick disk stars extending to lower metallicities
will be presented elsewhere.} thick disk stars that we observed 
are shown together with the 36 thick disk and 60 thin disk stars from
\cite{bensby2003, bensby2005} in a Toomre diagram in 
Figs.~\ref{fig:contour}d.

\section{Observations, data reduction, abundance analysis}

High-resolution ($R\approx65\,000$), high-quality ($S/N\gtrsim250$)
echelle spectra were obtained for $\sim 200$ thick disk F and G dwarfs by TB in 
Jan, Apr, and Aug in 2006 with the MIKE spectrograph \citep{bernstein2003} 
on the Magellan Clay 6.5\,m telescope at the Las Campanas Observatory in Chile.
Here we present the results for the metal-rich part of this sample, 
91 stars.
Solar spectra were obtained during the runs by observing the asteroid Vesta
(in Jan), the Jovian moon Ganymede (in Apr), and the asteroid Ceres (in Aug).

For the abundance analysis, we used the Uppsala MARCS stellar model
atmospheres \citep{gustafsson1975,edvardsson1993,asplund1997}.
The chemical compositions of the models were scaled with metallicity
relative to the standard solar abundances as given in
\cite{asplundgrevessesauval2005}, but with $\alpha$-element 
enhancements for stars with $\rm [Fe/H]<0$.
To determine the effective temperature and the microturbulence we required
all Fe\,{\sc i} lines to yield the same abundance, independent of lower
excitation potential, and line strength, respectively.
For the surface gravities, we utilized the accurate
{\sl Hipparcos} parallaxes \citep{esa1997} for our stars. Final abundances were
normalized on a line-by-line basis with our solar values as reference, and
then averaged for each element.

\section{Results}
\label{sec:results}

\begin{figure}
\plotone{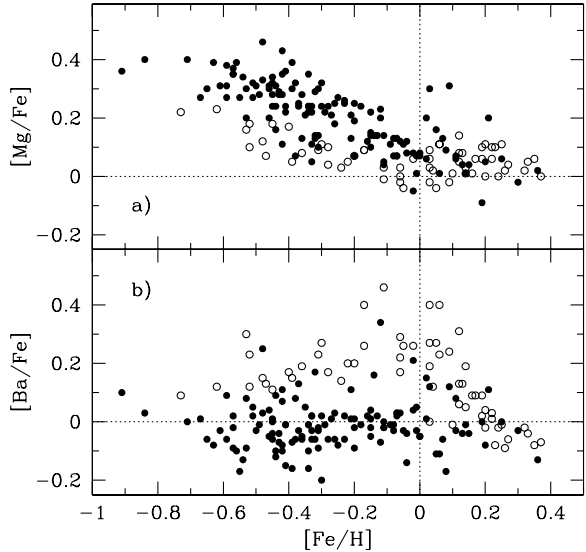}
\caption{
        [Mg/Fe] and [Ba/Fe] versus [Fe/H]. 
        Stars with thin and thick disk kinematics are marked
        by open and filled circles, respectively.
}
\label{fig:haltplottar0}
\end{figure}

\begin{figure}
\plottwo{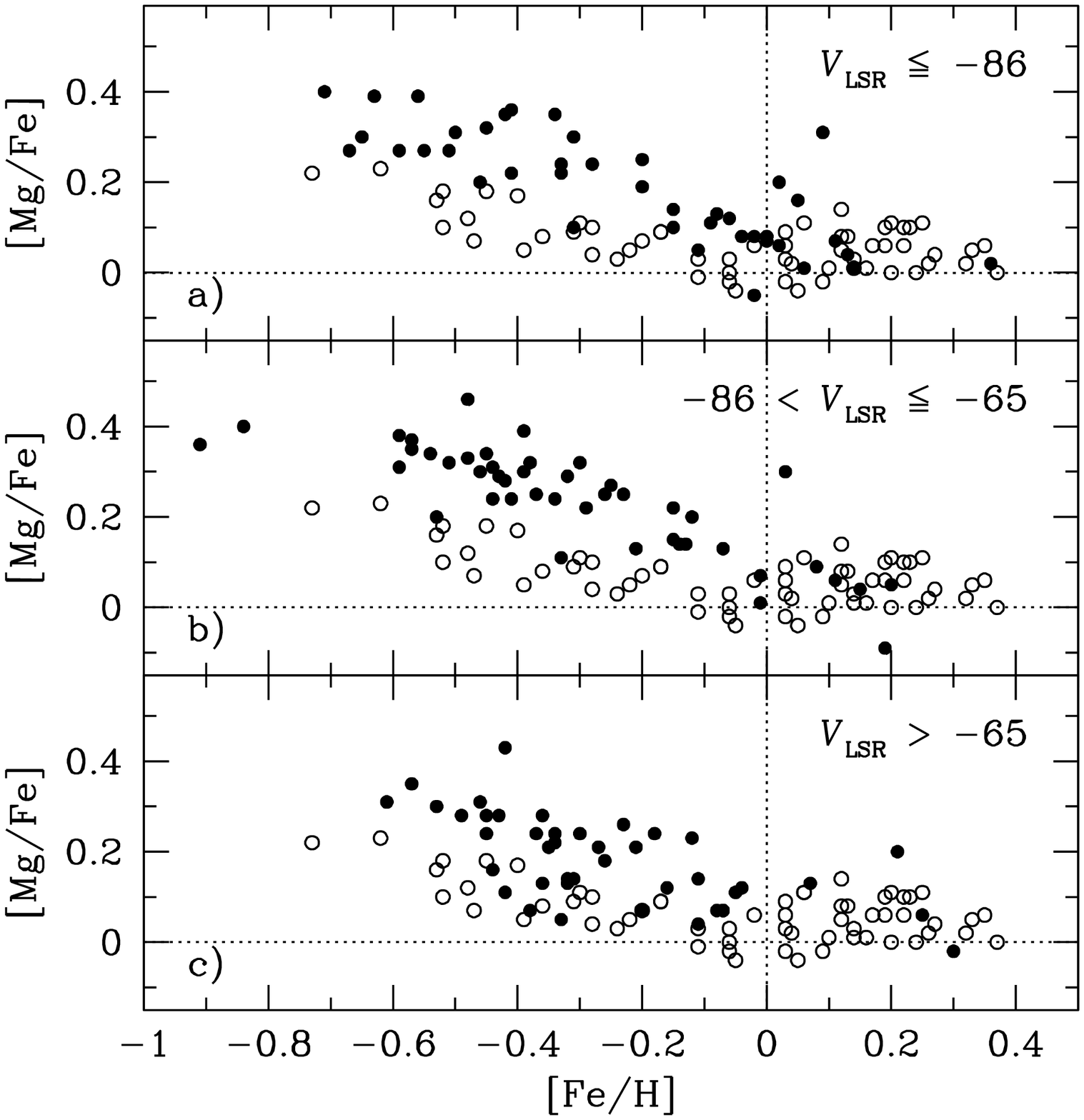}{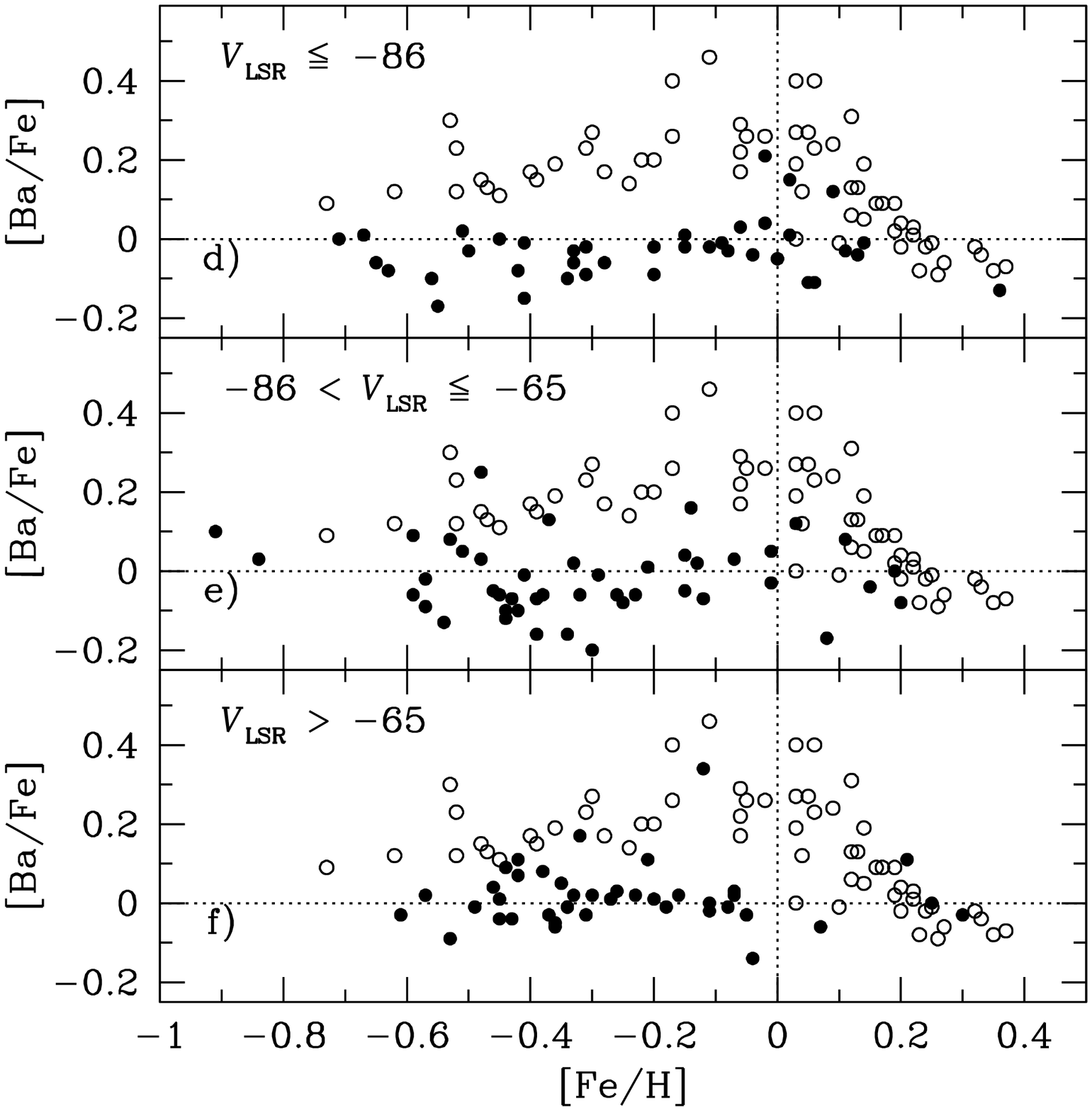}
\caption{
	[Mg/Fe] and [Ba/Fe] versus [Fe/H]. 
	The stars with thick disk kinematics have been
	split into three subsamples that span different $\vlsr$ velocities;
	$\vlsr\leq-86$, $-86<\vlsr\leq-65$, and $\vlsr>-65\,\kms$,
	respectively. These three regions are also marked in 
	Fig.~\ref{fig:contour}d.
	Stars with thin and thick disk kinematics are marked
	by open and filled circles, respectively.
}
\label{fig:haltplottar4}
\end{figure}

In Fig.~\ref{fig:haltplottar0} we show the resulting
$\rm [Mg/Fe]-[Fe/H]$ and $\rm [Ba/Fe]-[Fe/H]$ abundance trends for all
127 stars with thick disk kinematics and for the 60 stars
with thin disk kinematics.  It is clear that these two groups of stars
separate into distinct loci, and that those stars 
that can be associated with the thick disk show a flat [Mg/Fe] plateau
for metallicities below $\rm [Fe/H] \lesssim -0.4$, indicative of fast
enrichment from massive stars. Towards higher metallicities,
the [Mg/Fe] ratio for these stars declines, an indication of the delayed
enrichment from low- and intermediate mass stars. Approaching solar
metallicity, it also becomes harder to distinguish these stars from the
kinematically cold stars of the thin disk as their [Mg/Fe] trends merge.
The [Ba/Fe] ratio, on the other hand,
evolves essentially in lockstep with [Fe/H] for the kinematically hot
stars and is even more distinct from the thin disk [Ba/Fe]
trend, especially when the two disks approach solar metallicities.

In Fig.~\ref{fig:haltplottar0}, it is evident that a number of 
stars with thick disk kinematics have thin disk abundance ratios. 
In order to investigate the nature of these stars, we have 
divided the thick disk sample into three,
equally sized subsamples with different $\vlsr$ velocities (see also
Fig.~\ref{fig:contour}d): one that has low Galactic rotation velocities
($\vlsr\leq-86$, Figs.~\ref{fig:haltplottar4}a and d),
one that has intermediate rotation velocities
($-86<\vlsr\leq-65$, Figs.~\ref{fig:haltplottar4}b and e),
and one that has more thin-disk-like rotation velocities (but high
$\ulsr$ and/or $\wlsr$)
($\vlsr>-65\,\kms$, Figs.~\ref{fig:haltplottar4}c and f).

Since the sub-sample with $\vlsr\leq-86\,\kms$ deviates the most from the
typical thin disk kinematics, it should be the 
sample that is the least contaminated by the thin disk, and hence the most 
representative for the thick disk.
In Figs.~\ref{fig:haltplottar4}a and d,
we see that essentially all these stars are clearly distinguished from the
thin disk sample and have thick disk abundance
ratios. It is also clear that they indeed extend all the way up to solar
metallicity. The subsample with intermediate 
$\vlsr$ velocities ($-86<\vlsr\leq-65\,\kms$) 
is similar to the one with $\vlsr\leq-86\,\kms$
velocities, with the exception that a few  stars 
start to fall within the thin disk abundance trends.
The mixing of the populations is further increased in the
$\vlsr>-65\,\kms$ subsample, but still with most thick disk stars
clearly differentiated from the thin disk trend. In all $\vlsr$ bins
it is clearly demonstrated that the stars with thick disk kinematics and 
thick disk abundance ratios extend to at least $\rm [Fe/H]\approx0$. 

The gradual inclusion of stars
with thin disk abundance ratios as we approach higher $\vlsr$ velocities
is probably caused by the inclusion of more and more
stars from the high-velocity tail of the thin disk.

We have also investigated  the effects of similar cuts
in the $\ulsr$ and $\wlsr$ velocities for the thick disk sample,
and the results remain.

\section{Conclusion}

We have used high-resolution spectroscopy to trace the Galactic thick disk
to its highest metallicities. We find that the kinematically hot stars 
associated with the Galactic
thick disk extend to at least solar metallicity, if not above. 
We further note that our kinematic selection criteria
are not definitive, but tend to include a few stars 
from the high-velocity tail of the thin disk, in the thick disk sample,
especially at thin-disk--like $\vlsr$ velocities. 
In an upcoming paper, we will investigate  in detail
improvements to the kinematic selection criteria, to determine the
feasibility of weeding out the high-velocity tail of the thin disk
from kinematically selected thick disk samples.

\acknowledgements 

This work is supported by the National Science
Foundation, grant AST-0448900. SF is a Royal Swedish Academy
of Sciences Research Fellow supported by a grant from the Knut and
Alice Wallenberg Foundation.





\end{document}